\documentclass[conference]{IEEEtran}
\IEEEoverridecommandlockouts

\usepackage{cite}
\usepackage{amsmath,amssymb,amsfonts}
\usepackage{algorithm}
\usepackage{algorithmic}
\usepackage{graphicx}
\usepackage{textcomp}
\usepackage{xcolor}
\usepackage{url}
\usepackage{hyperref}
\usepackage{booktabs}
\usepackage{multirow}
\usepackage{float}
\usepackage{array}
\usepackage{verbatim}

\def\BibTeX{{\rm B\kern-.05em{\sc i\kern-.025em b}\kern-.08em
    T\kern-.1667em\lower.7ex\hbox{E}\kern-.125emX}}

\begin{document}

\makeatletter
\newcommand{\linebreakand}{%
  \end{@IEEEauthorhalign}
  \hfill\mbox{}\par
  \mbox{}\hfill\begin{@IEEEauthorhalign}
}
\makeatother

\title{PhishSnap: Image-Based Phishing Detection Using Perceptual Hashing}

\author{
\IEEEauthorblockN{Md. Abdul Ahad Minhaz}
\IEEEauthorblockA{\textit{Dept. of CSE} \\
\textit{United International University} \\
Dhaka, Bangladesh \\
mminhaz213072@bscse.uiu.ac.bd}
\and
\IEEEauthorblockN{Zannatul Zahan Meem}
\IEEEauthorblockA{\textit{Dept. of CSE} \\
\textit{United International University} \\
Dhaka, Bangladesh \\
zmeem213178@bscse.uiu.ac.bd}
\and
\IEEEauthorblockN{Dr. Md. Shohrab Hossain}
\IEEEauthorblockA{\textit{Professor, Dept. of CSE} \\
\textit{United International University} \\
Dhaka, Bangladesh \\
shohrab@cse.uiu.ac.bd}
}

\maketitle

\begin{abstract}
Phishing remains one of the most prevalent online threats, exploiting human trust to harvest sensitive credentials. Existing URL- and HTML-based detection systems struggle against obfuscation and visual deception. This paper presents \textbf{PhishSnap}, a privacy-preserving, on-device phishing detection system leveraging perceptual hashing (pHash). Implemented as a browser extension, PhishSnap captures webpage screenshots, computes visual hashes, and compares them against legitimate templates to identify visually similar phishing attempts. A \textbf{2024 dataset of 10,000 URLs} (70\%/20\%/10\% train/validation/test) was collected from PhishTank and Netcraft. Due to security takedowns, a subset of phishing pages was unavailable, reducing dataset diversity. The system achieved \textbf{0.79 accuracy}, \textbf{0.76 precision}, and \textbf{0.78 recall}, showing that visual similarity remains a viable anti-phishing measure. The entire inference process occurs locally, ensuring user privacy and minimal latency.
\end{abstract}

\begin{IEEEkeywords}
Phishing detection, perceptual hashing, browser extension, image similarity, cybersecurity, dataset bias.
\end{IEEEkeywords}

\section{Introduction}
Phishing attacks continue to dominate digital threat landscapes, responsible for a large proportion of credential theft incidents reported worldwide \cite{googleSafe}. Attackers host deceptive websites that visually imitate trusted services, exploiting human perception rather than technical vulnerabilities. The scale of phishing is illustrated by the Anti-Phishing Working Group (APWG), which recorded over 5 million phishing sites in 2024.

Traditional phishing detection mechanisms rely primarily on URL blacklists, lexical rules, or HTML-based pattern matching. These approaches are highly dependent on textual features, which adversaries can easily disguise through encoding, redirection, or domain obfuscation. Blacklist systems also fail to detect newly launched phishing sites that exist for only a few hours.

Machine learning-based approaches such as URLNet \cite{urlnet2018} and URLTran \cite{urltran2020} improve resilience by automatically learning URL representations. However, these models rely on textual data and large backend infrastructures. Hybrid systems that combine URL and HTML features \cite{hybrid2014,verma2017natural} perform better but remain limited when phishing pages depend primarily on visual imitation.

Human users often rely on visual cues—logos, layout, and color—to identify legitimate pages. Visual similarity-based approaches such as VisualPhishNet \cite{visualphishnet2019} and PhishZoo \cite{phishzoo2011} exploit this property. Yet, many of these models depend on deep CNNs hosted on remote servers, raising latency and privacy concerns.

\textbf{PhishSnap} introduces a new perspective by applying perceptual hashing (pHash) to phishing detection directly in the browser. The system computes a 64-bit visual signature of the current webpage and compares it to legitimate login templates locally. This ensures that no sensitive data ever leaves the user’s device. PhishSnap was trained and evaluated using a 2024 dataset of 10,000 URLs. However, as many phishing sites were already blocked or removed, the resulting screenshots were fewer, contributing to moderate accuracy scores.

\section{Related Work}
\subsection{URL and Lexical-Based Methods}
Le et al. proposed URLNet \cite{urlnet2018}, combining CNN and LSTM architectures to learn semantic representations of URLs. Zhang et al. introduced URLTran \cite{urltran2020}, employing transformers to improve URL context learning. While effective on large datasets, both are vulnerable to image-heavy or visually disguised phishing sites.

\subsection{Hybrid Content Approaches}
Banu and Saini \cite{hybrid2014} developed hybrid feature models integrating URL and HTML properties for classification. Verma and Das \cite{verma2017natural} demonstrated that rapid lexical and structural feature extraction improves scalability in real-time systems. These methods remain limited by their dependency on textual attributes.

\subsection{Visual Detection Techniques}
PhishZoo \cite{phishzoo2011} and VisualPhishNet \cite{visualphishnet2019} explored image-based similarity detection. PhishZoo used perceptual hashing to match webpage screenshots, while VisualPhishNet used deep CNNs to extract image embeddings. Earlier works on perceptual hashing, such as Monga and Evans \cite{monga2006perceptual} and Zauner \cite{zauner2010implementation}, established its robustness in detecting image tampering and visual duplicates. PhishSnap adopts these foundations to design a lightweight, on-device phishing detector requiring no cloud infrastructure.

\section{Methodology}
\subsection{Dataset and Splits}
A dataset of 10,000 URLs was compiled from PhishTank \cite{phishtank}, Netcraft \cite{netcraft}, and Alexa Top Sites in early 2024. The URLs were divided into 70\% for training, 20\% for validation, and 10\% for testing. Many phishing URLs were blocked or removed by security services such as Google Safe Browsing \cite{googleSafe}, resulting in missing screenshots and dataset imbalance across brands.

\subsection{Image Preprocessing}
Webpages were rendered at a resolution of 1366$\times$768 pixels to simulate a standard desktop environment. Screenshots were converted to grayscale and normalized to maintain uniform structure before hashing. This ensured that color variations did not affect visual similarity detection.

\subsection{Perceptual Hashing}
Perceptual hashing (pHash) captures the essential visual structure of an image. The process involves resizing the image to 32$\times$32, applying a Discrete Cosine Transform (DCT), extracting the low-frequency $8\times8$ coefficients, and converting each coefficient to a bit (1 if above mean, 0 otherwise). The resulting 64-bit hash serves as a compact fingerprint. Two images are compared using Hamming distance:
\[
d = \sum_{i=1}^{64} (H_1[i] \oplus H_2[i])
\]
Smaller distances indicate greater similarity. Threshold $T$ is chosen empirically based on validation results.

\begin{figure}[H]
\centering
\includegraphics[width=0.95\linewidth]{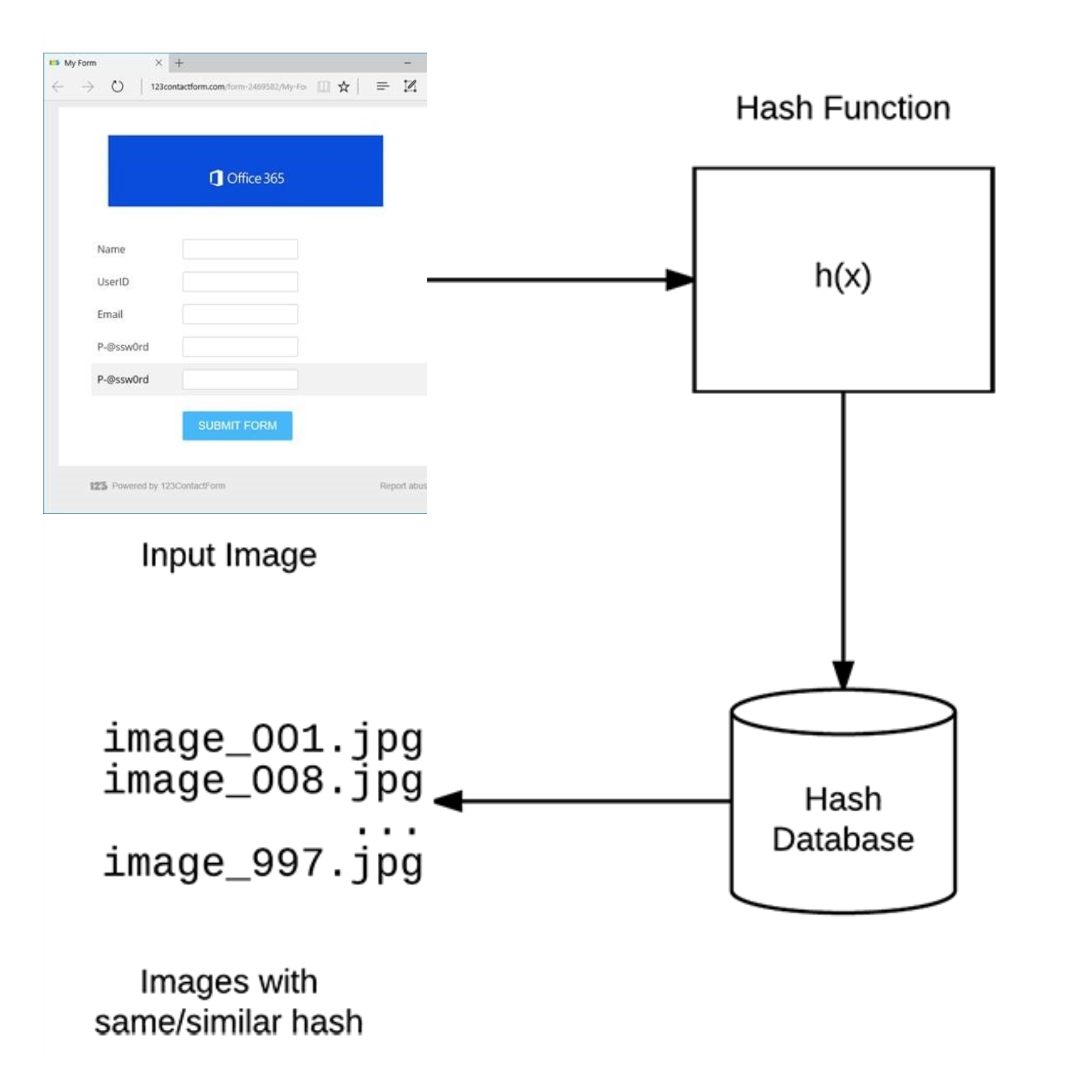}
\caption{Overview of the perceptual hashing pipeline: (a) normalized webpage capture; (b) grayscale conversion; (c) DCT and coefficient selection; (d) hash computation.}
\label{fig:phash}
\end{figure}

\subsection{Reference Hash Bank}
PhishSnap maintains a local reference JSON file containing precomputed pHashes of legitimate login pages from well-known brands. When a new page is scanned, its hash is compared against all stored templates. The minimum distance determines the classification:
\[
\text{Label} = 
\begin{cases}
\text{Phishing}, & d_{min} > T \\
\text{Safe}, & d_{min} \le T
\end{cases}
\]

\section{System Architecture}
PhishSnap operates entirely within the user’s browser using three components:
\begin{itemize}
    \item \textbf{Content Script:} Injected into the active tab to interface with the webpage.
    \item \textbf{Background Script:} Handles communication, screenshot capture, and computation requests.
    \item \textbf{Popup UI:} Provides an interface for users to trigger scanning and view results.
\end{itemize}

\begin{figure}[H]
\centering
\includegraphics[width=0.95\linewidth]{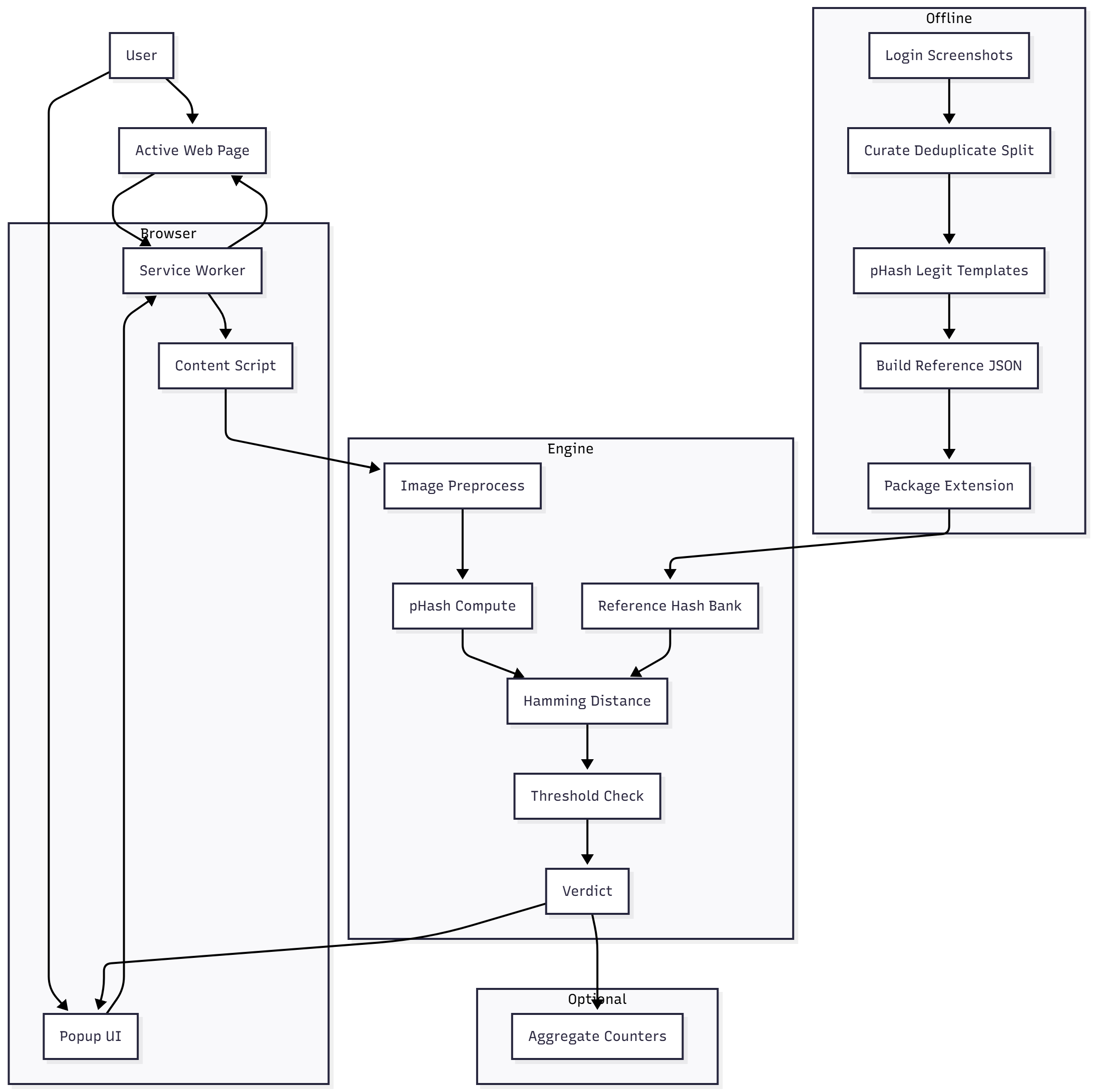}
\caption{System architecture of PhishSnap with browser integration and on-device processing.}
\label{fig:arch}
\end{figure}

All logic executes locally, ensuring that no screenshots or hashes leave the device. This approach reduces latency and enhances user privacy.

\section{Experimental Evaluation}
\subsection{Setup}
The model was evaluated on the held-out test set derived from the 2024 dataset. Each captured webpage was compared against a bank of 500 legitimate pHashes across multiple brands. Accuracy, precision, recall, and F1-score were calculated.

\begin{table}[H]
\centering
\caption{Evaluation Results on Test Set}
\begin{tabular}{@{}lc@{}}
\toprule
Metric & Value \\ \midrule
Precision & 0.76 \\
Recall & 0.78 \\
F1-score & 0.82 \\
Accuracy & 0.79 \\ \bottomrule
\end{tabular}
\end{table}

\subsection{Analysis}
The results confirm that perceptual hashing can distinguish legitimate and phishing pages based on visual similarity. Missing screenshots from blocked phishing URLs likely reduced the dataset’s variability, causing lower accuracy compared to ideal conditions. Nonetheless, PhishSnap’s on-device processing achieved reliable detection in under one second per scan.

\section{Extension Implementation}
The browser extension was implemented using HTML, CSS, and JavaScript with the Chrome Extension Manifest V3 API. Its workflow is divided into three scripts:
\begin{itemize}
    \item \textbf{Popup (popup.html):} Displays the “Scan Page” button and result text.
    \item \textbf{Background (background.js):} Captures the screenshot using \texttt{chrome.tabs.captureVisibleTab()}.
    \item \textbf{Content Script (content.js):} Performs pHash computation and compares results with stored JSON hashes.
\end{itemize}

\begin{figure}[H]
\centering
\includegraphics[width=0.9\linewidth]{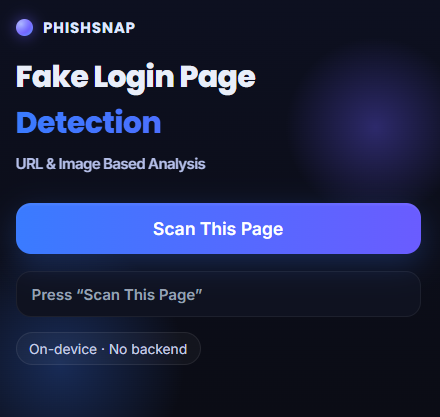}
\caption{PhishSnap browser extension showing the user interface and detection verdict.}
\label{fig:extension}
\end{figure}

The popup instantly displays “Safe” or “Phishing” along with a confidence score. All computations occur on the user’s machine; no network requests are made, ensuring privacy and speed.

\section{Limitations}
The dataset used for this study suffered from limited phishing diversity due to active takedowns during 2024. Pages that returned HTTP errors or security warnings could not be captured, leading to reduced visual representation. PhishSnap may also misclassify brandless phishing templates or adaptive designs that differ significantly from legitimate layouts. Future research will integrate textual analysis with perceptual hashing to enhance robustness.

\section{Conclusion}
PhishSnap demonstrates that perceptual hashing is an efficient and privacy-friendly approach to phishing detection. It eliminates the need for remote computation while maintaining strong detection performance. Future versions will explore hybrid visual-textual embeddings, adaptive thresholds, and support for mobile browsers to improve scalability and generalization.

\section*{Acknowledgment}
The authors thank the Department of CSE, United International University, for their resources, supervision, and guidance during this research.

\bibliographystyle{IEEEtran}
\bibliography{references}

\end{document}